\documentclass[conference]{IEEEtran}

\usepackage[cmex10]{amsmath}
\usepackage[utf8]{inputenc}
\usepackage{ucs, cite, amsfonts, amssymb, bm, bbm, graphicx, siunitx, color}
\usepackage[american]{babel}
\usepackage[T1]{fontenc}

\usepackage{algorithmic, algorithm}

\DeclareMathOperator{\diag}{diag}

\newcommand{\e}{\mathrm{e}}

\newcommand*{\herm}{^{\mathsf{H}}}
\newcommand*{\transp}{^{\mathsf{T}}}

\usepackage{fancyhdr, url, hyperref}
\hypersetup
{   
 colorlinks,
 citecolor=black,
 filecolor=black,
 linkcolor=black,
 urlcolor=blue,
}

\title{Beampattern design for radars\\ with reconfigurable intelligent surfaces}

\author{\IEEEauthorblockN{Emanuele~Grossi and Luca~Venturino}
\IEEEauthorblockA{\textit{Department of Electric and Information Engineering} \\
\textit{University of Cassino and Southern Lazio}\\
Cassino, Italy \\
e.grossi@unicas.it, l.venturino@unicas.it}
\thanks{This work was supported by the MIUR program ``Dipartimenti di Eccellenza 2018–2022.'' The authors are also with Consorzio Nazionale Interuniversitario per le Telecomunicazioni (CNIT), Parma, Italy.}
}

\begin{document}
\bstctlcite{BSTcontrol}

\IEEEoverridecommandlockouts
\maketitle

\thispagestyle{fancy}

\begin{abstract}
We consider a radar architecture where an illuminator composed of few sources is used as a feeder for a (passive) reconfigurable intelligent surface (RIS), so as to mimic the behavior of a multiple-input multiple-output (MIMO) radar composed of as many active elements as the RIS. In this framework, we study the problem of beampattern design in the space-frequency domain, and we propose to choose the source signals and the RIS adjustable phases in order to minimize the weighted squared error between the desired (amplitude) beampattern and the synthesized one. A low complexity iterative algorithm is proposed to solve the resulting non-convex least square problem. An example is provided to show the merits of the proposed approach.
\end{abstract}

\begin{IEEEkeywords}
Radar, Beampattern design, reconfigurable intelligent surface (RIS), multiple-input multiple-output (MIMO) radar.
\end{IEEEkeywords}

\section{Introduction}
 
A reconfigurable intelligent surface (RIS) is a low-cost passive flat surface made of sub-wavelength refractive/reflective elements that can add a tunable phase shift to the incident electromagnetic wave~\cite{Sha_2018, Huang_2019, DiRenzo_2020}. The elements of the RIS are controlled using an embedded logic with a (usually) negligible power consumption and can be used to redirect the incident wave in several ways, such as diffuse scattering, anomalous reflection, and beam focusing. The RIS can be used to control the radio propagation environment, thus providing novel degrees of freedom for the design of wireless systems~\cite{liu2021reconfigurable, Prasad2021, Jamali_2021}. RISs have been used to boost the performance of wireless communication links~\cite{Geoffrey-Ye-2020, Pan-2021, Basar-2021}; they have been also proven effective in other contexts, including wireless power transfer~\cite{zhao2020wireless}, localization and mapping~\cite{2020-Wymeersch-Localization-and-Mapping, Alouini-Localization}, and joint communication and sensing~\cite{joint_waveform}. 

\begin{figure}[t]	
\centering	 \centerline{\includegraphics[width=\columnwidth]{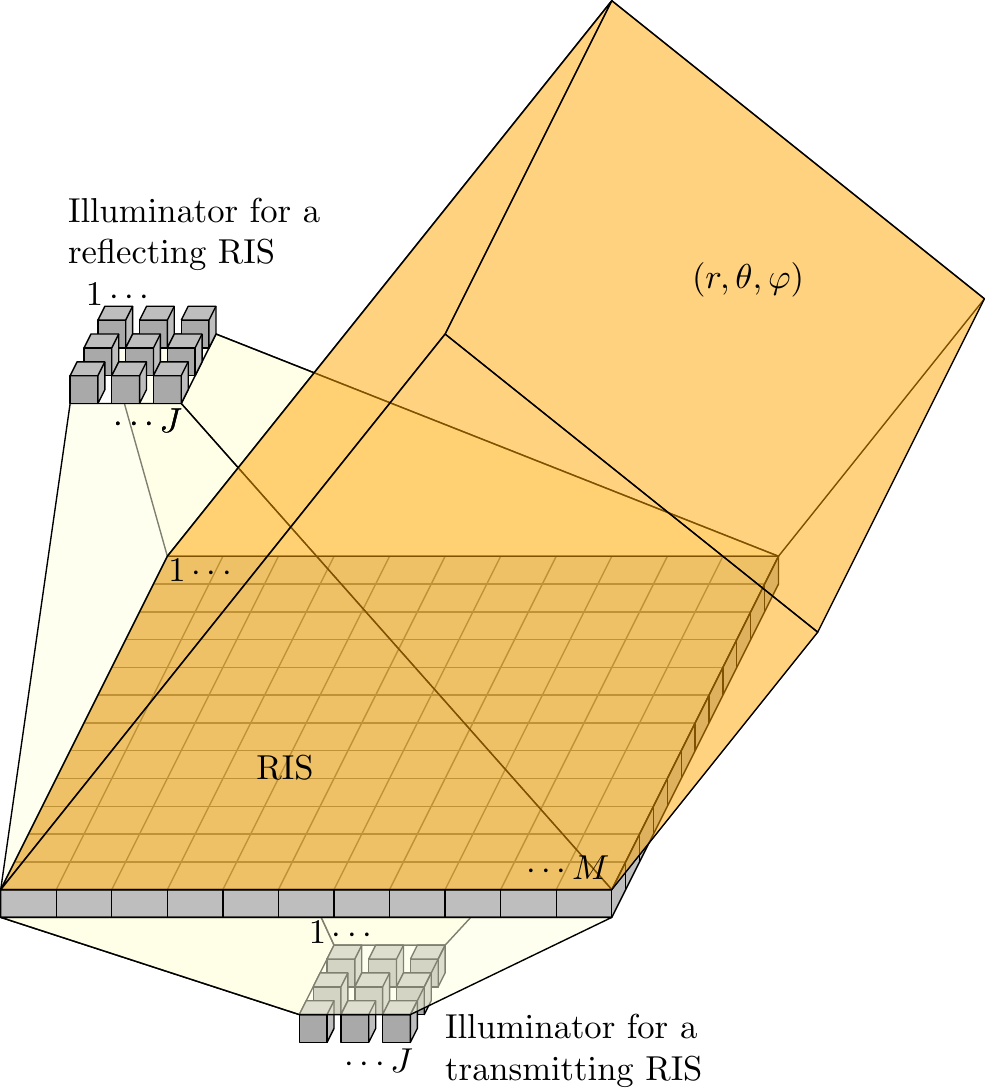}}	
 \caption{Considered radar architecture, composed of an illuminator with $J$ elements and a passive RIS with $M$ elements.} \label{fig_1}
\end{figure}

More recently, RISs have been shown to be advantageous also in enhancing the detection
capabilities of a radar system~\cite{Grossi2021ris, Aubry-2021, foundations, Grossi2022ris}, both in a widely-spaced and in a closely-spaced configuration. In particular, in~\cite{Grossi2021ris, foundations}, the idea of employing a low-cost radar sensor paired with a large RIS has been also advocated. In this configuration, shown in Fig.~\ref{fig_1}, the whole system acts as a feed antenna~\cite{Payam_2015}, were one or more feeders illuminate a large RIS, that is able to steer and focus the beam by properly modifying the tunable phase shifts.

In this work, we focus on the radar architecture shown in~Fig.~\ref{fig_1}, and we tackle the problem of beampattern design. This problem has already been addressed in the multiple-input multiple-output (MIMO) radar literature, for both wideband and narrowband cases~\cite{Stoica_2007, He_2011, Aubry_2016, Cheng_2017}, but, to the best of author's knowledge, it has not been explored in radar systems exploiting RISs. In the next section, we derive the expression of the beampattern in the far-field region for the proposed radar architecture in~Fig.~\ref{fig_1}. In Sec.~\ref{sys_opt_sec}, we formulate the beampattern design problem in the space-frequency domain as a least square problem: specifically, we propose to minimize the weighted squared error between the desired beampattern and the synthesized one over the source signals and the RIS phases with a constraint on the available power, and we offer a sub-optimal solution. In Sec.~\ref{num_res_sec}, we provide an illustrative example with a $2\times 2$ illuminator and a $10\times 10$ RIS, and we show that a satisfactory beampattern can be synthesized, also compared with a $10\times10$ MIMO radar, that features 100 active sources. Finally, concluding remarks are given in Sec.~\ref{concl_sec}.

\section{System description}  \label{sys_model_sec}

Here we consider a radar system composed of an illuminator with $J$ elements (also called sources or feeders) and a passive RIS with $M$ elements, as shown in Fig.~\ref{fig_1}. The illuminator is located on the same side as the target for a reflecting RIS and on the other side for a transmitting RIS (i.e., behind the RIS itself). In both cases, the distance between the $i$-th element of the RIS and the $j$-th source is $d_{ij}$. The signal emitted by source $j$ has carrier frequency $f_c$, support $[0,T]$, and bandwidth approximately equal to $W$; its low-pass equivalent signal is denoted $s_j(t)$. The phase shift introduced by element $i$ of the RIS, lying in position $\bm p_i$, is represented by the unit-modulus complex number $x_i$. We denote by $H_{ij}(f; r,\theta,\varphi)$ the frequency response of the channel linking the source $j$, the element $i$ of the RIS, and the observation point in the far-field region $(r, \theta, \varphi)$, and we model it as
\begin{equation}
 H_{ij}(f; r,\theta,\varphi) = x_i G_{ij}(f; \theta,\varphi) \frac{\e^{-2\pi i f ((d_{ij}+r)/c +\tau_i(\theta,\varphi))}}{4\pi d_{ij}r} \label{channel_expr}
\end{equation}
where $c$ denotes the speed of light, $4\pi d_{ij} r$ is the term due to the free-space attenuation, $(d_{ij}+r)/c +\tau_i(\theta,\varphi)$ is the propagation delay, with~\cite{Van_Trees_4}
\begin{equation}
 \tau_i(\theta,\varphi) = -\frac{1}{c} \left(p_{i,1}\cos\theta \cos\varphi +p_{i,2} \cos \theta \sin \varphi + p_{i,3}\sin \theta\right) \label{delays_expr}
\end{equation}
and $G_{ij}(f; \theta,\varphi)$ accounts for all the other effects of the channel.

With this notation, the Fourier transform of the lowpass equivalent signal observed from the point $(r,\theta,\varphi)$ is
\begin{align}
 Y(f;r,\theta,\varphi) &= \sum_{i=1}^M \sum_{j=1}^J S_j(f) H_{ij}(f+f_c; r,\theta,\varphi)\notag\\
 &= \frac{1}{\sqrt{4\pi} r} \e^{-2\pi i (f+f_c)r/c} \sum_{i=1}^M \e^{-2\pi i (f+f_c) \tau_i(\theta,\varphi)} x_i \notag\\
 & \quad \times \sum_{j=1}^J \frac{\e^{-2\pi i (f+f_c)d_{ij}/c}}{\sqrt{4\pi} d_{ij}} G_{ij}(f+f_c; \theta,\varphi) S_j(f) \label{Y_expr}
\end{align}
where $S_j(f)$ is the Fourier transform of $s_j(t)$. We now let $\bm x  = [ x_1 \; \cdots \; x_M]\transp \in \mathbb C^M$, $\bm \sigma(f) = [S_1(f) \; \cdots \; S_J(f)] \transp \in \mathbb C^J$, and $\tilde{\bm G}(f; \theta,\varphi) \in \mathbb C^{M\times J}$ with entry $(i,j)$ defined as
\begin{equation}
 \tilde G_{ij}(f; \theta,\varphi) =\frac{1}{\sqrt{4\pi}d_{ij}} \e^{-2\pi i (f+f_c)d_{ij}/c} G_{ij}(f+f_c; \theta,\varphi). \label{G_tilde_expr}
\end{equation}
Then,~\eqref{Y_expr} can also be written as
\begin{multline}
 Y(f;r,\theta,\varphi)  = \frac{1}{\sqrt{4\pi} r}\e^{-2\pi i (f+f_c)r/c}\\
\times \bm v\herm(f;\theta,\varphi) \diag \bm x \tilde{\bm G}(f; \theta,\varphi) \bm \sigma(f)
\end{multline}
where $\bm v(f;\theta,\varphi) = [\e^{2\pi i (f+f_c)\tau_1(\theta,\varphi)} \; \cdots \; \e^{2\pi i (f+f_c)\tau_M(\theta,\varphi)} ]\transp$ is the steering vector in the direction $(\theta,\varphi)$ at frequency $f$~\cite{Van_Trees_4}, and $\diag \bm x$ is a diagonal matrix with the elements of $\bm x$ on the main diagonal; the (amplitude) beampattern at frequency $f$ and spatial angle $(\theta, \varphi)$ is, therefore,
\begin{align}
 B(f; \theta,\varphi) &= \sqrt{\frac{4\pi r^2}{T}}\bigl| Y(f;r,\theta,\varphi)\bigr|\notag\\
 &=\frac{1}{\sqrt T} \bigl| \bm v\herm(f;\theta,\varphi) \diag \bm x \tilde{\bm G}(f; \theta,\varphi) \bm \sigma(f) \bigr|. \label{ampl_beampattern}
\end{align}

We now exploit the fact that the signals are bandlimited (as commonly done; e.g., see~\cite{He_2011}); in particular, we can write
\begin{equation}
 S_j(f)=\int_0^T s_j(t) \e^{-2\pi i ft}dt \approx \frac{1}{W} \sum_{n=1}^N s_j(n/W) \e^{-2\pi nf/W}
\end{equation}
where $N=\lfloor WT\rfloor$. Therefore, upon defining $\bm s_n = [ s_1 (n/W)$ $ \cdots \; s_J(n/W)]\transp \in \mathbb C^J$, and $\bm s = [\bm s_1\transp \; \cdots \; \bm s_N\transp ]\transp \in \mathbb C^{JN}$, we can approximate~\eqref{ampl_beampattern} as
\begin{align}
 & B(f;\theta,\varphi)  \notag\\
 & \quad \approx \frac{1}{W\sqrt T} \biggl| \bm v\herm(f;\theta,\varphi) \diag \bm x \tilde{\bm G}(f; \theta,\varphi) \sum_{n=1}^N \bm s_n \e^{-2\pi i nf/W} \biggr|
 \notag\\
 &\quad = \frac{1}{W \sqrt T} \Bigl| \bm v\herm(f;\theta,\varphi) \diag \bm x \tilde{\bm G}(f; \theta,\varphi) \bigl( \bm e\transp(f) \otimes \bm I_J \bigr) \bm s \Bigr|\notag\\
 &\quad =  \bigl| \bm v\herm(f;\theta,\varphi) \diag \bm x \bm Q (f;\theta, \varphi) \bm s \bigr|
\end{align}
where
\begin{align}
 \bm e(f)&= \begin{bmatrix}\e^{-2\pi i f/W} & \cdots & \e^{-2\pi i N f /W} \end{bmatrix}\transp \in \mathbb C^{N}\\
 \bm Q (f; \theta, \varphi)& = \frac{1}{W \sqrt T} \tilde{\bm G}(f; \theta,\varphi) \bigl( \bm e\transp(f) \otimes \bm I_J \bigr) \in \mathbb C^{M\times JN}.
\end{align}
In the following, we tackle the problem of beampattern design.

\section{System Optimization} \label{sys_opt_sec}

The objective is to design the signal emitted by the illuminator $\bm s$ and the RIS adjustable phases $\bm x$ in such a way that the amplitude beampattern $B(f;\theta, \varphi)$ matches the desired beampattern, say $d(f; \theta, \varphi)$. To this end, we discretize the angular region $[-\pi/2, \pi/2]^2$ with $L$ points, namely $\{ (\theta_\ell, \varphi_\ell)\}_{\ell=1}^L$, and the frequency region $[-W/2, W/2]$ with $K$ points, namely $\{f_k\}_{k=1}^K$, and we use a least square approach for the design. The beampattern matching problem tackled here is, therefore,
\begin{equation}
 \begin{aligned}
 \min_{\bm s \in \mathbb C^{JN}, \bm x \in \mathbb C^M } & \; \sum_{k=1}^K \sum_{\ell=1}^L w_{k\ell} \bigl( d(f_k;\theta_\ell, \varphi_\ell) - B(f_k;\theta_\ell, \varphi_\ell)\bigr)^2\\
 \text{s.t.} & \; \frac{1}{N} \Vert \bm s \Vert^2 \leq P \\
 & \; |x_i|=1, \quad \forall i
 \end{aligned} \label{problem_1}
\end{equation}
where $\{w_{k\ell}\}_{k\ell}$ are given weights and $P$ is the available power. This design criterion has been extensively used in the past years (see, e.g.,~\cite{He_2011}); the semidefinite (instead of definite) constraint for the power budget is needed since there are two hops here (source-RIS and RIS-destination), and the same beampattern level can be synthesized in different ways, since the source signals can combine constructively/destructively on the RIS. For notational convenience, we let $d_{k\ell}= d(f_k;\theta_\ell,\varphi_\ell)$, $\bm Q_{k\ell} = \bm Q(f_k;\theta_\ell,\varphi_\ell)$, and $\bm v_{k\ell}= \bm v(f_k;\theta_\ell,\varphi_\ell)$, for $k=1,\ldots,K$ and $\ell=1,\ldots,L$, so that  Problem~\eqref{problem_1} can be equivalently rewritten as
\begin{equation}
 \begin{aligned}
 \min_{\bm s \in \mathbb C^{JN}, \bm x \in \mathbb C^M} & \; \sum_{k=1}^K \sum_{\ell=1}^L w_{k\ell} \bigl( d_{k\ell} - | \bm v_{k\ell}\herm \diag \bm x \bm Q_{k\ell} \bm s | \bigr)^2\\
 \text{s.t.} & \; \Vert \bm s \Vert^2 \leq N P \\
 & \; |x_i|=1, \quad \forall i.
 \end{aligned}\label{problem_2}
\end{equation}

The objective function of Problem~\eqref{problem_2} is non-differentiable, and, as commonly done~\cite{He_2011}, we exploit the fact that, for any $a>0$ and $b\in\mathbb C$,
\begin{equation}
 \min_{\psi \in \mathbb R} | a \e^{i\psi} - b |^2 = \bigl( a - | b| \bigr)^2, \text{ for } \psi=\arg b \label{trick}
\end{equation}
where $\arg$ is the argument (or phase) of a complex number. Thus, introducing the auxiliary variables $\{\psi_{k\ell}\}_{k\ell}$, Problem~\eqref{problem_2} can be rewritten as
\begin{equation}
 \begin{aligned}
 \min_{\substack{\bm s \in \mathbb C^{JN}, \bm x \in \mathbb C^M,\\ \{\psi_{k\ell}\}_{k\ell}\in \mathbb R^{KL}}} & \; \sum_{k=1}^K \sum_{\ell=1}^L w_{k\ell} \left| d_{k\ell}\e^{i\psi_{k\ell}} - \bm v_{k\ell}\herm \diag \bm x \bm Q_{k\ell} \bm s\right|^2\\
 \text{s.t.} & \; \Vert \bm s \Vert^2 \leq N P \\
 & \; |x_i|=1, \quad \forall i.
 \end{aligned} \label{problem_3}
\end{equation}
This problem is non-convex, and we tackle it by resorting to the block-coordinate descent method~\cite{Bertsekas_1999}, also known as non-linear Gauss-Seidel method or as alternating minimization: starting from a feasible point, the objective function is minimized with respect to each of the ``block coordinate'' variables, taken in cyclic order, while keeping the other ones fixed at their previous values. Here, the block-coordinate variables are the auxiliary variables $\{\psi_{k\ell}\}_{k\ell}$, the illuminator signals $\bm s$, and the RIS adjustable phases $\bm x$, and the corresponding three reduced-complexity sub-problems are addressed in the remainder of this section.

\subsection{Minimization over the auxiliary variables}\label{min_psi_sec}

The variables $\{\psi_{k\ell}\}_{k\ell}$ can be disjointly minimized, and, exploiting~\eqref{trick}, we have that the solution is
\begin{equation}
\psi_{k\ell}=\arg (\bm v_{k\ell}\herm \diag \bm x \bm Q_{k\ell} \bm s)
\end{equation}
for $k=1,\ldots,K$ and $\ell=1,\ldots,L$.


\subsection{Minimization over the illuminator signals}\label{min_s_sec}

We define here
\begin{subequations} \label{def_A_b_q}
\begin{align}
 \bm A & = \sum_{k=1}^K \sum_{\ell=1}^L w_{k\ell} \bm Q_{k\ell}\herm \diag \bm x^*\bm v_{k\ell} \bm v_{k\ell}\herm \diag \bm x \bm Q_{k\ell} \label{mat_A_def}\\
 \bm b &= \sum_{k=1}^K \sum_{\ell=1}^L w_{k\ell} d_{k\ell} \e^{i \psi_{k\ell}} \bm Q_{k\ell}\herm \diag \bm x^* \bm v_{k\ell} 
\end{align}%
\end{subequations}
so that the sub-problem to be solved can be rewritten as
\begin{equation}
 \min_{\substack{\bm s \in \mathbb C^{JN}:\\ \Vert\bm s \Vert^2 \leq N P}} \bigl\{ \bm s\herm \bm A \bm s - 2 \Re (\bm s\herm \bm b) \bigr\} \label{sub_prob_s_2}
\end{equation}
where $\Re (\,\cdot\,)$ is the real part. This is a convex problem, were a quadratic function is to be minimized over a ball, and the solution can be found by imposing the Karun-Kush-Tucker conditions. A closed-form expression (up to the Lagrangian multiplier, that can be found through a simple bisection algorithm) for the solution is available.

\subsection{Minimization over the RIS phases}\label{min_x_sec}

Notice that $\diag \bm x \bm Q_{k\ell} \bm s = \diag (\bm Q_{k\ell} \bm s) \bm x$, and then the problem is
\begin{equation}
 \min_{\substack{\bm x \in \mathbb C^M: \\ |x_i|=1 \forall i}} \sum_{k=1}^K \sum_{\ell=1}^L w_{k\ell} \left| d_{k\ell}\e^{i\psi_{k\ell}} - \bm v_{k\ell}\herm \diag (\bm Q_{k\ell} \bm s) \bm x \right|^2  \label{sub_prob_x_1}
\end{equation}
If we define
\begin{equation}
 \bm B = \sum_{k=1}^K \sum_{\ell=1}^L w_{k\ell} \left[ \begin{matrix}
 \bm c_{k\ell} \bm c_{k\ell}\herm & - d_{k\ell} \e^{i \psi_{k\ell}} \bm c_{k\ell}\\
 - d_{k\ell} \e^{-i \psi_{k\ell}} \bm c_{k\ell}\herm & d_{k\ell}^2\\
 \end{matrix}\right]
\end{equation}
where $\bm c_{k\ell} =  \diag (\bm Q_{k\ell} \bm s)^* \bm v_{k\ell}$, Problem~\eqref{sub_prob_x_1} can be re-written as
\begin{equation}
 \min_{\substack{\bm x \in \mathbb C^M: \\ |x_i|=1 \forall i}} \begin{bmatrix}\bm x\herm & 1\end{bmatrix} \bm B \begin{bmatrix}\bm x\transp & 1\end{bmatrix}\transp \label{sub_prob_x_2}
\end{equation}
that, as it can be easily verified, is equivalent to
\begin{equation}
 \min_{\substack{\bm z \in \mathbb C^{M+1}: \\ |z_i|=1 \forall i}} \bm z\herm \bm B\bm z .\label{sub_prob_x_4}
\end{equation}
If $\bm z^\star$ solves~\eqref{sub_prob_x_4}, the solution to Problem~\eqref{sub_prob_x_2} can be recovered as $x_i^\star=z_i^\star/z_{M+1}^\star$, $i=1,\ldots, M$.

Problem~\eqref{sub_prob_x_4} is a NP-hard complex quadratic problem~\cite{Zhang_Huang_2006, Luo_2010}. A suboptimal solution can be found by using the coordinate descent algorithm, where each entry of $\bm z$ is iteratively minimized: in this case, we have
\begin{equation}
 z_i= \begin{cases} -\sum_{j\neq i} B_{ij} z_j, & \text{if } \sum_{j\neq i} B_{ij} z_j\neq 0\\
 \text{any } z: |z|=1, & \text{otherwise.}
\end{cases}
\end{equation}
Alternatively, the problem can be tackled by resorting to the projected gradient method, or by reformulating it as a trace minimization problem that can sub-optimally be solved with a semi-definite relaxation followed by a Gaussian randomization~\cite{Goemans_2004, Zhang_Huang_2006, Luo_2010,Waldspurger_2015}.

\section{Numerical results} \label{num_res_sec}

We consider here a $10 \times 10$ transmitting RIS, so that $M=100$, illuminated by $J=4$ sources. 
The signals have a duration $T=\SI{0.64}{\us}$ and a bandwidth approximately equal to $W=\SI{100}{\MHz}$; the carrier frequency is $f_c=\SI{3}{\GHz}$. The inter-element spacing of the RIS is half-wavelength, while the 4 feeders are placed at a distance of \SI{60}{\cm} from the RIS, each one in correspondence with one of the four quadrants of the RIS. The number of frequency sampling points is $K=64$, and they are uniformly spaced in $[-\SI{50}{\MHz}, \SI{50}{\MHz}]$; in the angular domain, $36 \times 36$ sampling points uniformly spaced in $[-90^\circ, 90^\circ]\times [-90^\circ, 90^\circ]$ are taken, so that $L=1296$. The desired beampattern is composed of two beams with equal height: a (broad) beam in the region $[-45^\circ,0^\circ] \times [-45^\circ, 0^\circ,]$ for the negative frequencies and a (narrow) beam in region $[22.5^\circ,45^\circ] \times [22.5^\circ,45^\circ]$ for all frequencies.  The available power is $P=\SI{10}{\W}$, the number of signal samples is $N=\lfloor WT\rfloor=64$, and the weights are set equal to one for all $k$ and $\ell$. In all figures, we plot the power beampattern normalized by its maximum value, i.e.,
\begin{equation}
  \text{NPB}(f_k;\theta_\ell,\varphi_\ell) = \frac{B^2(f_k;\theta_\ell,\varphi_\ell)}{\max_{(k', \theta', \varphi')} B^2(f_{k'};\theta_{\ell'},\varphi_{\ell'})}.
 \end{equation}
Also, for the beampattern matching problem, we report the relative square error, i.e.,
\begin{equation}
 \text{RSE}=\frac{\sum_{k=1}^K \sum_{\ell=1}^L w_{k\ell} \bigl( d(f_k;\theta_\ell, \varphi_\ell) - B(f_k;\theta_\ell, \varphi_\ell)\bigr)^2}{\sum_{k=1}^K \sum_{\ell=1}^L w_{k\ell} d^2(f_k;\theta_\ell, \varphi_\ell)}.
\end{equation}
Finally, for the sake of comparison, we also consider the case of a $10 \times 10$ MIMO radar.

\begin{figure}[t]	
\centering
\centerline{\includegraphics[width=\columnwidth]{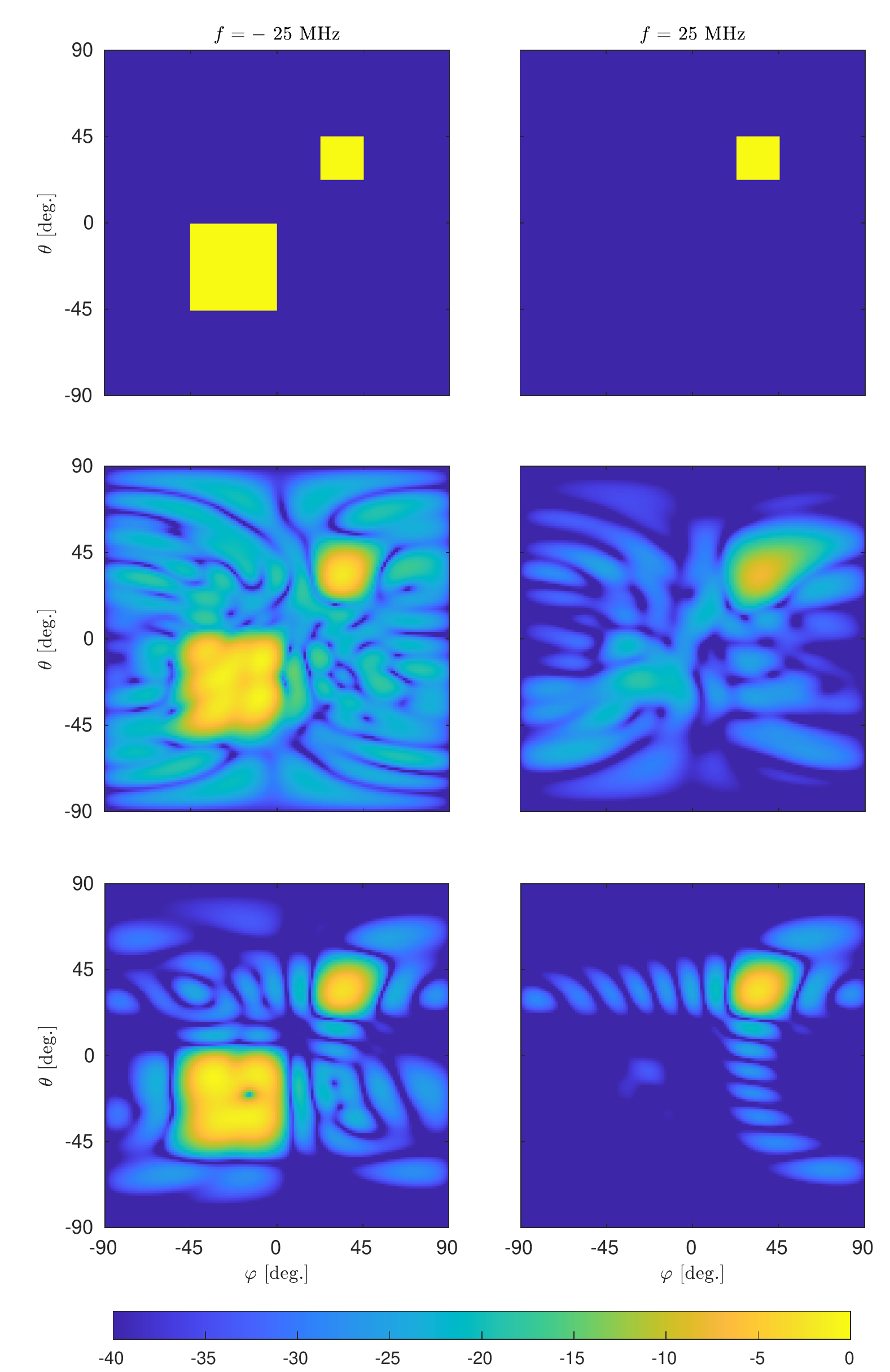}}	
 \caption{Desired (top row), synthesized with the considered radar architecture (middle row), and synthesized with a MIMO radar (bottom row) normalized power beampattern in dB scale as a function of elevation and azimuth for $-25$ MHz (left column) and 25 MHz (right column).} \label{fig_2}
\end{figure}

\begin{figure}[t]	
\centering	 \centerline{\includegraphics[width=\columnwidth]{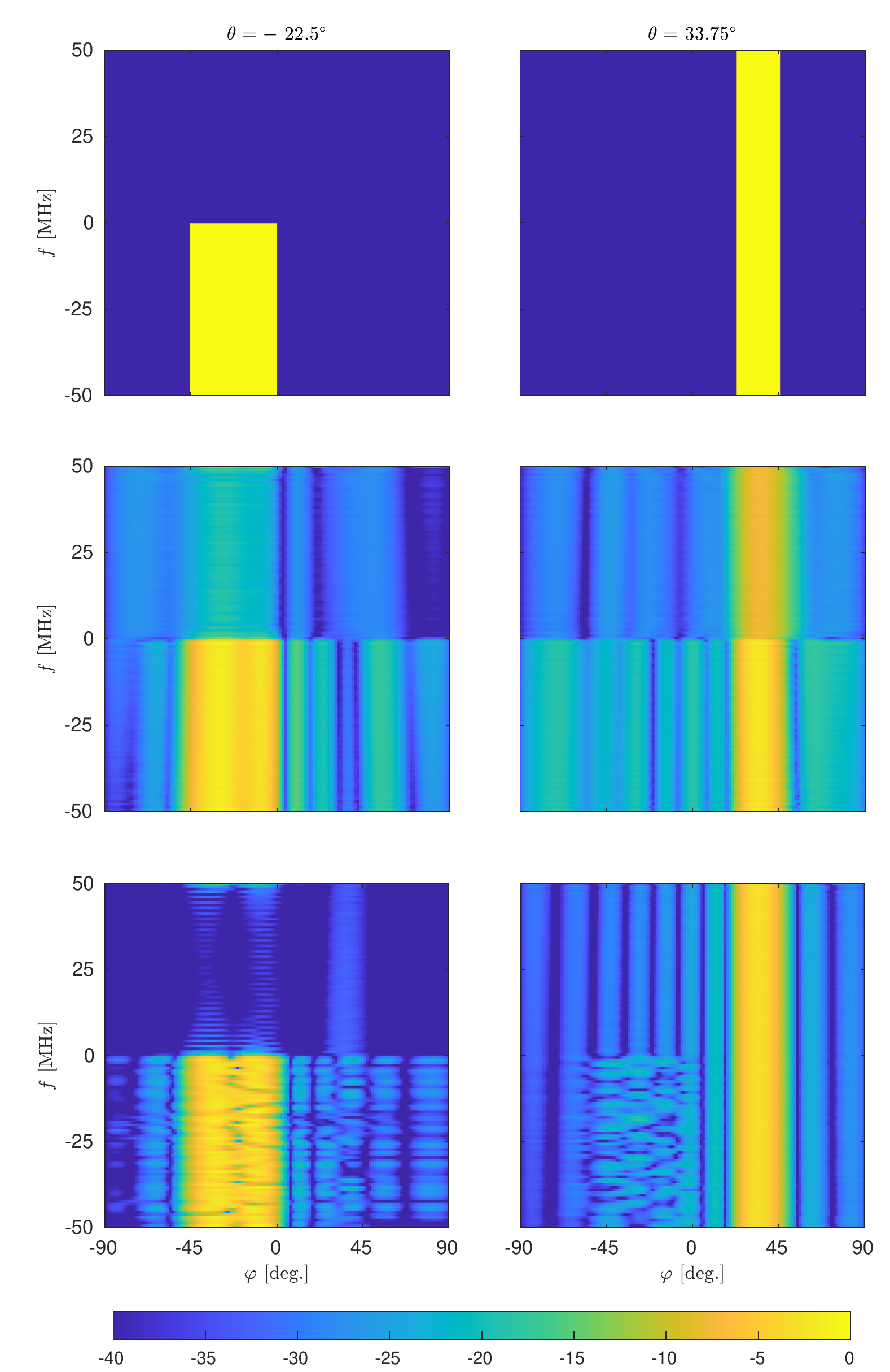}}	
 \caption{Desired (top row), synthesized with the considered radar architecture (middle row), and synthesized with a MIMO radar (bottom row) normalized power beampattern in dB scale as a function of frequency and azimuth for elevations $-22.5^\circ$ (left column) and $33.75^\circ$ (right column).} \label{fig_3}
\end{figure}

\begin{figure}[t]	
\centering	 \centerline{\includegraphics[width=\columnwidth]{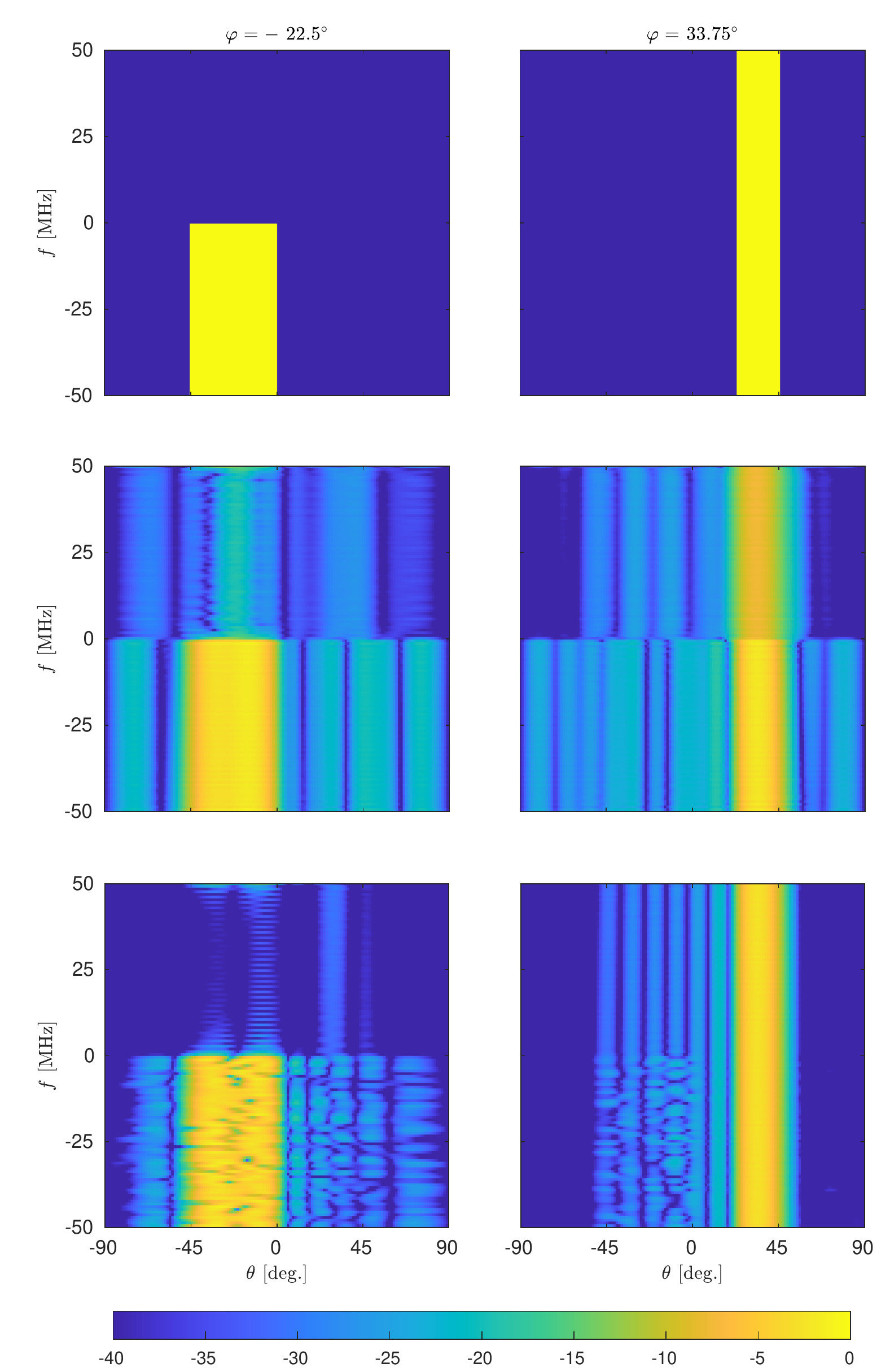}}	
 \caption{Desired (top row), synthesized with the considered radar architecture (middle row), and synthesized with a MIMO radar (bottom row) normalized power beampattern in dB scale as a function of frequency and elevation for azimuths $-22.5^\circ$ (left column) and $33.75^\circ$ (right column).} \label{fig_4}
\end{figure}

In Fig.~\ref{fig_2}--\ref{fig_4}, we show (in dB scale) the desired NPB (top row) and the synthesized NPB for the architecture in Fig.~\ref{fig_1} (middle plot) and for the MIMO radar case (bottom plot). Specifically we plot the NPB as a function of azimuth and elevation for two different frequencies in Fig.~\ref{fig_2}, as a function of frequency and azimuth for two different elevations in Fig.~\ref{fig_3}, and as a function frequency and elevation for two different azimuths in Fig.~\ref{fig_4}. It can be seen from the figures that the beampattern synthesized with the architecture in Fig.~\ref{fig_1} is close to the desired one, and that it does not lose too much from the upper bound represented by the MIMO radar; in particular, the RSE is $0.21$ for the considered architecture and $0.14$ for the MIMO radar, that, however, employs a much larger number of active elements.

\section{Conclusion} \label{concl_sec}

We considered the radar architecture in Fig.~\ref{fig_1}, where an illuminator is used to feed a passive RIS, and we tackled the problem of beampattern design by optimizing the RIS adjustable phases and the source signals under a power constraint. The provided example has shown that a satisfactory performance can be achieved with a 100-element RIS and as few as 4 sources, and that the synthesized beampattern is very similar to the one realized with a 100-element MIMO radar. Future studies will include additional constraints on the source waveforms (such as the peak-to-sidelobe level) and will focus on the optimization of the number and position of the sources.

\bibliographystyle{IEEEtran}

\end{document}